\documentclass[11pt]{article}

\usepackage{epsf,latexsym,amssymb}
\usepackage{graphicx}
\usepackage{setspace}
\usepackage{natbib}
\usepackage{url}
\newcommand{\journal}[1]{}
\newcommand{\preprint}[1]{}

\newcommand{\KKK}{MAPKKK}
\newcommand{\KKKP}{MAPKKK-P}
\newcommand{\KK}{MAPKK}
\newcommand{\KKP}{MAPKK-P}
\newcommand{\KKPP}{MAPKK-PP}
\newcommand{\K}{MAPK}
\newcommand{\KP}{MAPK-P}
\newcommand{\KPP}{MAPK-PP}
\newcommand{\mKKK}{\mbox{[MAPKKK]}}
\newcommand{\mKKKP}{\mbox{[MAPKKK-P]}}
\newcommand{\mKK}{\mbox{[MAPKK]}}
\newcommand{\mKKP}{\mbox{[MAPKK-P]}}
\newcommand{\mKKPP}{\mbox{[MAPKK-PP]}}
\newcommand{\mK}{\mbox{[MAPK]}}
\newcommand{\mKP}{\mbox{[MAPK-P]}}
\newcommand{\mKPP}{\mbox{[MAPK-PP]}}
\newcommand{\Sig}{Sig}
\newcommand{\mSig}{\mbox{[Sig]}}

\newcommand{\mKm}{\mbox{Km}}
\newcommand{\Ki}{\Ki}
\newcommand{\mKi}{\mbox{K}_i}
\newcommand{\mV}{\mbox{V}}

\newcommand{\comment}[1]{}

\newcommand{\rref}[1]{(\ref{#1})}

\newtheorem{theorem}{Theorem}
\newtheorem{itlemma}{Lemma}[section] 
\newtheorem{itproposition}[itlemma]{Proposition}
\newtheorem{itcorollary}[itlemma]{Corollary}
\newtheorem{itremark}[itlemma]{Remark}
\newtheorem{itdefinition}[itlemma]{Definition}
\newtheorem{itexample}[itlemma]{Example}
\newenvironment{lemma}{\begin{itlemma}\rm}{\end{itlemma}} 
\newenvironment{remark}{\begin{itremark}\rm}{\end{itremark}} 

\newenvironment{proposition}{\begin{itproposition}\rm}{\end{itproposition}}
\newenvironment{definition}{\begin{itdefinition}\rm}{\end{itdefinition}}
\newenvironment{example}{\begin{itexample}\rm}{\end{itexample}}
\def\bi{\begin{itemize}}
\def\ei{\end{itemize}}
\def\ben{\begin{enumerate}}
\def\een{\end{enumerate}}
\def \beq {\begin{eqnarray}}
\def \eeq {\end{eqnarray}}
\def \beqn {\begin{eqnarray*}}
\def \eeqn {\end{eqnarray*}}

\newcommand{\text}[1]{\hbox{\rm \ #1\ \/}}
\newcommand{\be}[1]{\begin{equation}\label{#1}}
\newcommand{\ee}{\end{equation}}
\newcommand{\bl}[1]{\begin{lemma}\label{#1}}
\newcommand{\br}[1]{\begin{remark}\label{#1}}
\newcommand{\bt}[1]{\begin{theorem}\label{#1}}
\newcommand{\bd}[1]{\begin{definition}\label{#1}}
\newcommand{\bp}[1]{\begin{proposition}\label{#1}}
\newcommand{\bc}[1]{\begin{itcorollary}\label{#1}}
\newcommand{\ec}{\mybox\end{itcorollary}}
\newcommand{\ecs}{\end{itcorollary}}
\newcommand{\bfact}[1]{\begin{fact}\label{#1}}
\newcommand{\bex}[1]{\begin{example}\label{#1}}
\newcommand{\bem}[1]{\begin{example}\label{#1}}  
\newcommand{\efact}{\mybox\end{fact}}
\newcommand{\eex}{\mybox\end{example}}
\newcommand{\eem}{\mybox\end{example}}
\newcommand{\el}{\mybox\end{lemma}}
\newcommand{\ele}{\mybox\end{lemmaex}}
\newcommand{\er}{\mybox\end{remark}}
\newcommand{\et}{\qed\end{theorem}}
\newcommand{\ed}{\mybox\end{definition}}
\newcommand{\ep}{\mybox\end{proposition}}
\newcommand{\epr}{\end{proof}}
\newcommand{\bpr}{\begin{proof}}

\newcommand{\eers}{\end{exercise}}
\newcommand{\eexs}{\end{example}}
\newcommand{\eems}{\end{example}}
\newcommand{\els}{\end{lemma}}
\newcommand{\eles}{\end{lemmaex}}
\newcommand{\ers}{\end{remark}}
\newcommand{\ets}{\end{theorem}}
\newcommand{\eds}{\end{definition}}
\newcommand{\eps}{\end{proposition}}
\newcommand{\qed}{\hfill \halmos} 
\newcommand{\mybox}{\hfill $\Box$} 
\newcommand{\halmos}{\rule{1ex}{1.4ex}}
\newenvironment{proof}{\noindent {\em Proof}.\ }{\hspace*{\fill}$\halmos$\medskip}

\def\edo{\end{document}}

\comment{possible referees:
Hofmeyr, W'Hoff, Boris, Fell
}

\begin{document}

\title{MAPK Cascades as Feedback Amplifiers}

\comment{
\author{Herbert M Sauro$^{123}$ and Brian Ingalls$^{43}$ \\ \\ $^1$Department of Bioengineering,
University of Washington, Seattle, WA, 98195 \\ \\
$^4$Department of Applied Mathematics, University of Waterloo, \\
Waterloo, Ontario,
Canada N2L 3G1 \\ \\
\\
$^2$ Correspondence: Herbert M Sauro, Department of Bioengineering, University of Washington, Seattle \\
$^3$ This work was largely carried out while employed at the \\
California Institute of Technology during 2002. }}

\author{Herbert M Sauro$^{13}$ and  Brian Ingalls$^{23}$ \\ \\ $^1$
Corresponding Author: Department of Bioengineering, \\ University of Washington, Seattle, WA, 98195 \\ \\
$^2$Department of Applied Mathematics, University of Waterloo, \\
Waterloo, Ontario,
Canada N2L 3G1 \\
\\
$^3$ This work was largely carried out while employed at the \\
California Institute of Technology during 2002.
\\ \\
Running Head: Biological Feedback Amplifiers}

\date{}

\maketitle

\pagebreak

\begin{abstract}
\noindent Interconvertible enzyme cascades, exemplified by the
mitogen activated protein kinase (MAPK) cascade, are a frequent
mechanism in signal transduction pathways. There has been much
speculation as to the role of these pathways, and how their
structure is related to their function.  A common conclusion is that
the cascades serve to amplify biochemical signals so that a single
bound ligand molecule might produce a multitude of second
messengers. Some recent work has focused on a particular feature
present in some MAPK pathways -- a negative feedback loop which
spans the length of the cascade.  This is a feature that is shared
by a man-made engineering device, the feedback amplifier. We propose
a novel interpretation: that by wrapping a feedback loop around an
amplifier, these cascades may be acting as biochemical feedback
amplifiers which imparts i) increased robustness with respect to
internal perturbations; ii) a linear graded response over an
extended operating range; iii) insulation from external
perturbation, resulting in functional modularization. We also report
on the growing list of experimental evidence which supports a graded
response of MAPK with respect to Epidermal Growth Factor. This
evidence supports our hypothesis that in these circumstances MAPK
cascade, may be acting as a feedback amplifier.
\end{abstract}

\pagebreak

\section*{Introduction}

The ability of biological cells to receive and respond to signals,
such as environmental conditions, mating potential or developmental
cutes is considered a fundamental characteristic of life. In
multicellular organisms, cells will often receive a flood of signals
aimed at regulating their behavior. It is not surprising therefore
to discover that cells have evolved highly elaborate and complex
networks of `signaling' pathways whose role is to coordinate and
integrate this information and to elicit a suitable response. One of
the most surprising results to have come to light in the study of
these signaling pathways is the remarkable degree of evolutionary
conservation that exists among widely different
organisms~\citep{Marshall}. The family of MAPK (mitogen-activated
protein kinase) pathways, in particular, is highly conserved.

The MAPK pathways, besides being highly conserved, are common
components in signal transduction pathways \citep{Chang:2001}.
Virtually all eukaryotic cells that have been examined (ranging from
yeast to man) possess multiple MAPK pathways each of which responds
to multiple inputs. In mammalian systems MAPK pathways are activated
by a wide range of input signals including a variety of growth
factors and environmental stresses such as osmotic shock and
ischemic injury \citep{Kyriakis2001,MAPKBook2002}. Once the MAPK
pathways have integrated these signals, they coordinately activate
gene transcription with resulting changes in protein expression
leading to cell cycling, cell death and cell differentiation.

\comment{Because yeast genetic manipulation is well advanced, much
of the work on MAPK pathways has been done on budding yeast and
through this work the notion of multiple parallel MAPK signaling
pathways was developed. Currently, six {\em S.\ cerevisiae} MAPK
signaling pathways have been identified \citep{Banuett:1998}. Many
of the pathways receive multiple signals with overlapping  inputs
resulting in cross-talk between the input stages of the MAPK
pathways. The picture that is emerging is that of at least two
functional layers, a preprocessing layer responsible for receiving
the inputs and channelling them to one or more MAPK stages in a
second layer for final delivery to the target, namely
transcriptional modulation (see Figure 1).}

Modern methods of genetic manipulation have allowed extensive
investigation of MAPK pathways in budding yeast. Through this work
\citep{Marshall} the notion of multiple parallel MAPK signaling
pathways was developed. At least six {\em S.\ cerevisiae} MAPK
cascades have been identified \citep{Banuett:1998}.  Many of these
pathways receive multiple signals, some of which are shared with
other MAPK cascades.  This signalling architecture allows for
cross-talk between the input stages of the MAPK pathways.  The
emerging picture is that of signal transduction pathways consisting
of least two functional layers: a decision-making layer responsible
for receiving and integrating the inputs, which are then channeled
to one or more MAPK stages in a second layer for final delivery to
the target, in this case transcriptional modulation (see Figure 1).

\begin{figure}[h]
  \includegraphics{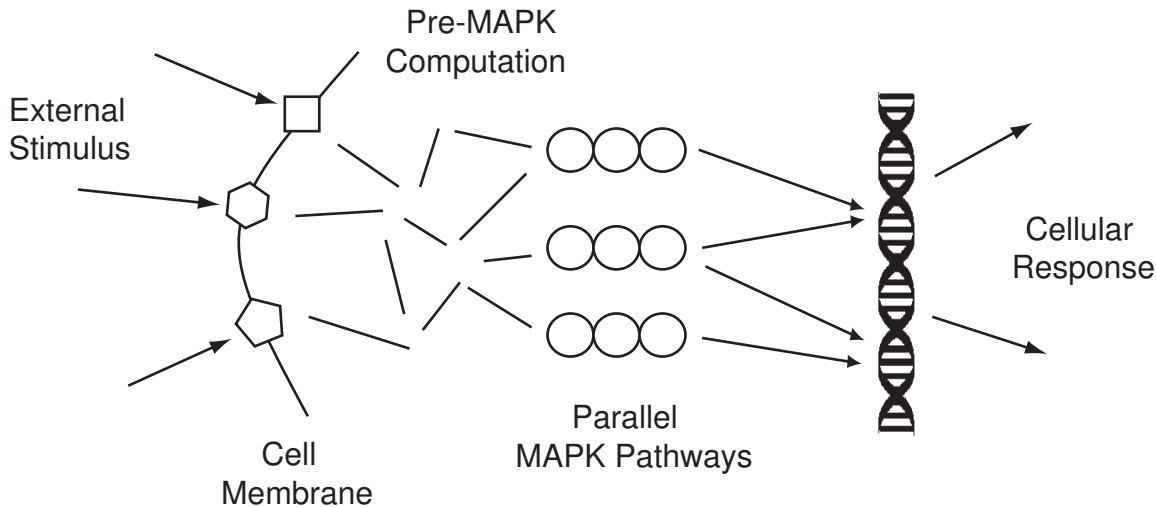}
  \caption{MAPK Pathways}\medskip
\end{figure}

In this paper we address the functional nature of the MAPK pathway
stage. What properties do MAPK pathways possess which might suggest
their role in this biochemical computation?  There are a number of
characteristics which are common to all MAPK families and have a
bearing on how we should answer this question. Firstly, the
structure of these pathways can give rise to extremely high gains on
the input/output response, of the order of hundreds of thousands in
some cases.  Secondly, MAPK cascades always appear as the final
stage of a signal transduction pathway.  To our knowledge there has
never been reported any further networks between the last stage of a
MAPK cascade and transcriptional modulation.  Finally, these
cascades appear to have little interaction with the rest of the
system beyond their basic input/output behavior \citep{Widmann1999}.
In fact, this functional separation may be achieved in part by a
physical separation, as some evidence suggests that the components
of MAPK cascades are held together by scaffolding proteins which may
provide a spatial separation from the rest of the signal
transduction network~\citep{Widmann1999,Garrington1999}.

These observations lead to a satisfying interpretation of MAPK
cascades as modular components of signal transduction pathways which
act to amplify their inputs. Others have also noted the possibility
that MAPK cascades can act as switch devices~\citep{Ferrell:1996},
or even delay systems~\citep{Nelson2004,Ihekwaba2005}. In this paper
we wish to investigate the hypothesis that MAPK pathways, may in
some circumstances be acting as feedback amplifiers.

The paper is organized as follows. The structure and function of
MAPK cascades is considered first.  The concept of Feedback
amplifiers is introduced next, where the analysis is kept simple by
considering linear systems. The benefits of feedback in the MAPK
cascade are illustrated in the next section, both by simulation
results and by analysis of the linearization of the model.  A
discussion of experimental evidence is included in the last section.

\begin{figure}[h]
\begin{center}
  \includegraphics[width=3.8in,height=3.4in]{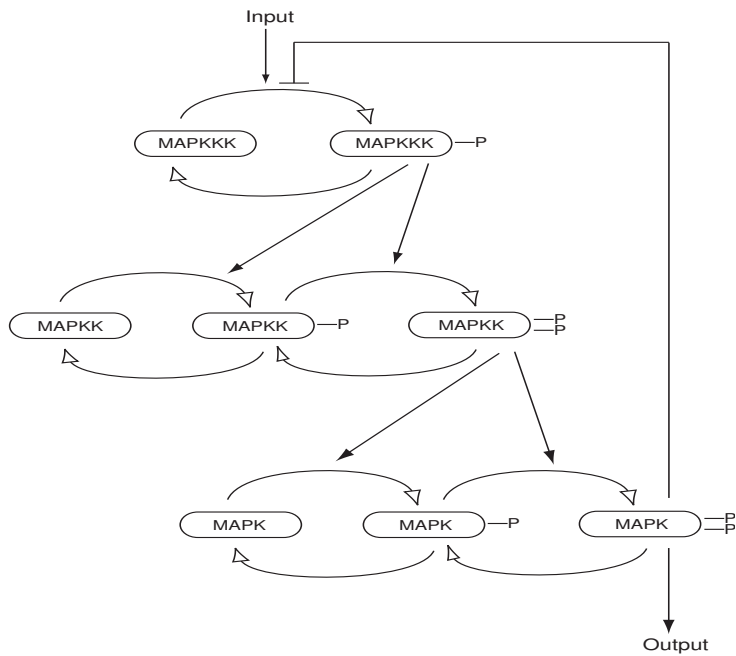}
  \caption{Schematic of a Generic MAPK Pathway}\medskip
\label{network}
\end{center}
\end{figure}

\paragraph{Structure and Properties of Cascades} As
discussed in the introduction, MAPK cascades are common components
of extracellular signal-regulated pathways. Experimental results
suggest that these cascades are highly conserved across a wide
spectrum of organisms~\citep{Marshall}. In addition, the basic
structure of the pathway in each MAPK family is essentially the
same. This implies that the functional roles played by each family
is the same, or very similar. The architecture involves three
protein kinases (see Figure 2), each of which activates the kinase
below only after being phosphorylated from above.

A compelling explanation of the function of these cascades is that
they provide a switch-like response
\citep{StadtmanChock:1977a,Goldbeter:1984} which may be used in
decision making, for example in development. It has been appreciated
for many years that even a single cycle can elicit switch-like
behavior under conditions in which the cycle reactions are operating
near saturation \citep{GDKosh81}. A more general analysis was made
by Small and Fell \citep{FSm86b,SF90a} and Cardenas and
Cornish-Bowden \citep{Cardenas:1989} using metabolic control
analysis and subsequently extended by Kahn and Westerhoff
\citep{Kahn:1991} and Kholodenko et al. \citep{BKBrown:1997}.
Arguments for switch-like behavior  are supported by experimental
evidence that some MAPK cascades may exhibit positive feedbacks
leading to bistable behavior.~\citep{Ferrell:2002,BhallaRam}

During a switching action, the `output' of the system (the
concentration of activated kinase at the end of the cascade \KPP) is
switched from its basal (non-stimulated) level to saturation
(i.e.~100\% activation), as the `input' (e.g.~ligand level) crosses
a threshold value~\citep{Ferrell:1996}. This behavior is often
described by the analogy to a cooperative enzyme -- the cascade acts
as a single enzyme with a large Hill coefficient.  Along with its
function as a memoryless switch, the cascade also amplifies the
input signal, since the pool of MAPK typically has a concentration
orders of magnitude higher than the ligand levels which activate the
response.

In addition studies have indicated the presence of one or more
negative feedback loops surrounding the MAPK
cascade~\citep{Downward1996}. This feedback is mediated by
phosphorylation of SOS by ERKPP which in turn results in
dissociation of the ShcGS complex thus interfering with Ras
activation of the ERK pathway.

In this paper we suggest a novel interpretation of the role of
negative feedback in a MAPK cascade, inspired by analogous
situations in electrical engineering. Amplifiers are some of the
most ubiquitous components in electrical circuits.  All amplifiers
necessarily saturate above and below certain input levels.  They are
designed so that their `active range' covers the levels of input of
interest. When this active range is small, the result is switch-like
behavior, suggesting an analogy with MAPK pathways.

While amplification of signals is a common function of electrical
circuits, amplifiers are rarely used in isolation.  Rather, negative
feedback is introduced from output to input, resulting in a {\em
feedback amplifier}. These devices represent the key component of
all (electrical) analog computation, since they allow the
construction of functional {\em operational amplifiers}. The
benefits of wrapping feedback around an amplifier may not be
intuitively obvious, and indeed came as a surprise to many people
when the idea was introduced. The primary improvements achieved with
feedback are i) increased robustness with respect to internal
perturbations; ii) a linear graded response over an extended
operating range; iii) insulation from external perturbation,
resulting in functional modularization.

In the electronics industry these properties are exploited in the
manufacture of amplifiers.  Cheap, low tolerance, high-gain
amplifier components are mass produced.  These amplifiers are not
suitable for use alone, but rather are coupled with high
tolerance passive resistor components which implement feedback
and thus significantly improve the characteristics of the
amplifier.

We present the hypothesis that evolution may have hit upon the
benefits of the feedback amplifier, and is implementing it in the
MAPK cascade. In doing so we build on the work of others who have
suggested that negative feedback may serve to improve the function
of biochemical amplifiers \citep{BhallaRam,Cinquin2002,Sa76}.

In its active range, the cascade would act solely as an amplifier --
providing a faithful amplification of the input presented to it from
the mechanisms upstream.  To these upstream mechanisms (which we do
not address) would then be relegated the all-important task of
integrating the myriad extracellular signals received by the cell.
The MAPK cascade then plays the role of amplifying the results of
those biochemical calculations to the point where they can produce a
response in the cell.  This is an obvious strategy, again
exemplified in electrical engineering. Analog computers perform
their calculations on tiny currents and voltages then, when
finished, amplify the results so that they may be useful. The same
energy-conserving principle may be used in the cell. Signals may be
processed at small concentrations (with correspondingly low needs
for biosynthesis of enzymes).  Only after the computation is
complete and the appropriate response has been determined is it
necessary to amplify concentrations to the point where they will
affect the activity of the cell.

That the MAPK pathway serves as the tail-end for wide variety of
different signaling pathways indicates that it may be an example of
{\em modular} design.  This design principle produces component
which can play their role in a number of different situations.  It
may be that the MAPK cascade serves as a ``plug-and-play'' amplifier
with specificity determined by inclusion of the appropriate
``adapter'', serving to connect the cascade to its input and output.
This interpretation is supported by the finding that the
``internal'' mechanism of the MAPK cascade (i.e.~the phosphorylation
of MAPKK and MAPK) are highly conserved while the ``connectors''
(phosphorylation of MAPKKK and the target of MAPK-PP) vary according
to the role of the cascade~\citep{Marshall,Garrington1999}

This conservation of structure and mechanism is a great boon to the
modeler.  Based on this evidence we can have some confidence that
analysis of a model of the cascade may provide insight into the
biochemistry of a wide variety of signal transduction pathways. With
that hope in mind, we present an analysis of a model of the
Raf/Ras/MEK/ERK pathway presented in~\citep{BKoholodenko:2000} and
based on work in~\citep{Ferrell:1996}. The model does not
include the important mechanisms which occur upstream from the
cascade.

Despite the evidence for conservation of the cascade, there are
variations in structure and mechanism which may or may not
correspond to differences in pathway response.  Common variants are
cascades which are longer or shorter than three kinases, or in which
the kinases are phosphorylated at only one site.  Previous work
suggests that such differences may not have much of an effect on the
function of the pathway~\citep{Brightman:2000,Goldbeter:1984}. A
more extreme variation is a cascade in which the activation is not
performed by kinases, but rather by methylation or acetylation.
There are fewer experimental results in such cases, but work by
modelers suggests that an analogous functionality would be expected.
In particular, in this paper we are concerned primarily with the
cascade's qualitative role as an amplifier, which will likely be
maintained over a wide variety of biochemical implementations.

\paragraph{Feedback Amplifiers} Rather than species concentrations,
the amplifiers used by electrical engineers are designed to amplify current
or voltage in an electrical circuit.  A simple example is the voltage
amplifier (i.e.~voltage controlled voltage source) which samples the
voltage in one part of a circuit and produces a proportional voltage in
another.

Amplification is one of the most fundamental tasks one can demand
of an electrical circuit. One of the challenges facing engineers
in the 1920's was how to design amplifiers whose performance was
robust with respect to the internal parameters of the system and
which could overcome inherent nonlinearities of their
implementation.  This problem was especially critical to the
effort to implement long distance telephone lines across the
U.S.A.

These difficulties were overcome by the introduction of the {\em
feedback amplifier}, designed in 1927 by Harold S. Black
\citep{Mindell}, who was an engineer for Western Electric (the
forerunner of Bell Labs). The basic idea was to introduce a negative
feedback loop from the output of the amplifier to its input. At
first sight, the addition of negative feedback to an amplifier might
seem counterproductive\footnote{As Horowitz and Hill put it
in~\citep{ArtOfElectronics} `Negative feedback is the
process of coupling the output back in such a way as to cancel some
of the input.  You might think that this would only have the effect
of reducing the amplifier's gain and would be a pretty stupid thing
to do.'}.  Indeed Black had to contend with just such opinions when
introducing the concept -- his director at Western Electric
dissuaded him from following up on the idea, and his patent
applications were at first dismissed.  In his own words, ``our
patent application was treated in the same manner as one for a
perpetual motion machine''~\citep{blackhist}.

While Black's detractors were correct in insisting that the negative
feedback would reduce the gain of the amplifier, they failed to
appreciate his key insight -- that the reduction in gain is
accompanied by increased robustness of the amplifier and improved
fidelity of signal transfer.  This trade-off between gain and
system performance can be elegantly demonstrated by considering linear
systems, to which we now turn.

\begin{figure}
\begin{center}
  \includegraphics[width=5.8in,height=3in]{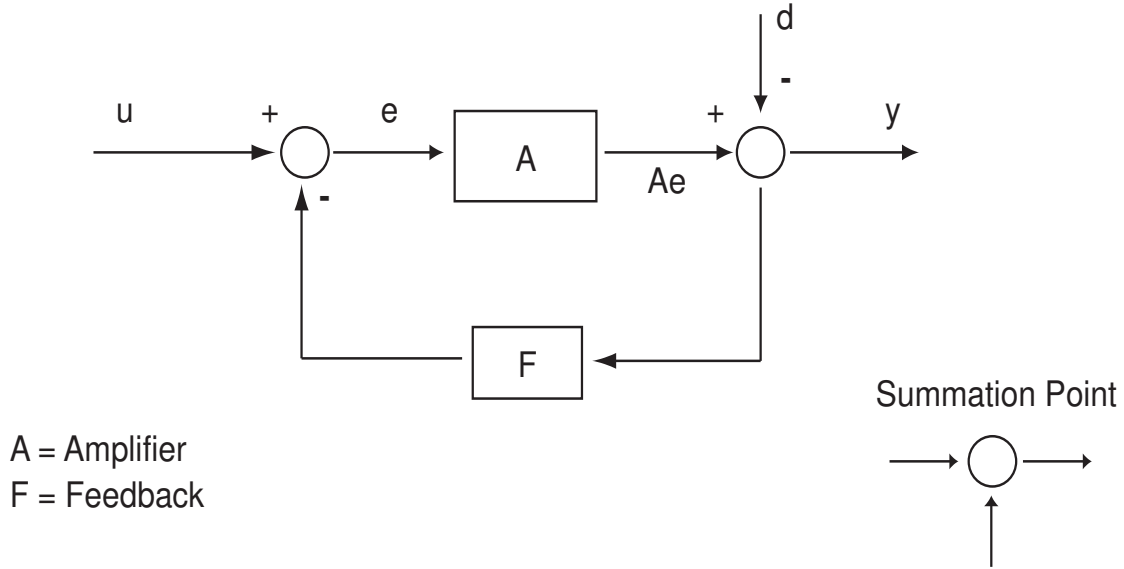}
\end{center}
\caption{Linear Feedback System: $d=$ disturbance; $u=$ input; $y=$ output}
\label{feedblock}
\end{figure}

\section*{Results}

Consider the block diagram in Figure~\ref{feedblock}. We will
consider only the steady-state behavior of the system, so we take
the input $u$, the output $y$, the error $e$, and the disturbance
$d$ to be constants.  To begin, consider the case of no disturbance
($d=0$). Assume that both the amplifier $A$ and the feedback $F$ act
by multiplication. Without feedback (i.e. with $F=0$), the system
behavior is described by $y = Au$, which is an amplifier with (open
loop) gain $A$.

Introducing feedback, the behavior of the system is as follows. From the
diagram \beqn y= Ae \qquad \mbox{and} \qquad e = u - Fy. \eeqn Eliminating
$e$, we find \beq \label{amp1} y = \frac{Au}{1+AF} \eeq Calling
$G=\frac{A}{1+AF}$ the {\em system} (or {\em closed loop}) {\em gain}, we
have simply $y=Gu$. Comparing $G$ with $A$, it is immediate that the
feedback does indeed reduce the gain of the amplifier.  Further, if the
{\em loop gain} $AF$ is large ($AF \gg 1$), then \beqn G \approx
\frac{A}{AF} = \frac{1}{F}. \eeqn That is, as the gain $AF$ increases, the
system behavior becomes more dependent on the feedback loop and less
dependent on the amplifier itself.  We next indicate three specific
consequences of this key insight.

\paragraph{Sensitivity to internal parameter variation} Since the
closed loop gain $G$ depends on the amplifier gain $A$, a small
change $\Delta A$ in the gain $A$ will produce a change $\Delta G$
in $G$. Considering the ratio of relative changes we find \beq
\label{amp2} \frac{\partial G}{\partial A} \frac{A}{G} =
\frac{1}{1+AF} \qquad \mbox{so} \qquad \frac{\Delta G}{G} \approx
\frac{1}{1+AF} \frac{\Delta A}{A} \eeq provided $\Delta A$ is small.
That is, a 1\% change in the open loop gain $A$ leads to a roughly
$\frac{1}{1+AF}$ \% change in the system gain.  We see that as the
loop gain $AF$ is increased the system becomes less sensitive to
perturbations in its internal structure.  In electrical circuit
design this might alleviate the problems caused by temperature
fluctuations or aging components.  In a biochemical network the
result will be a system whose function is not sensitive to
fluctuations in enzyme concentrations. It should be noted that this
reduction in sensitivity comes at a price -- as the loop gain is
increased the relative sensitivity to changes in $F$ grows to unity.
As mentioned, this performance constraint can be met by coupling low
tolerance amplifiers to carefully tuned feedback elements.  At this
point we can only speculate as to whether nature has hit upon the
same design strategy

\paragraph{Sensitivity to disturbances in the output} Suppose now that
a constant disturbance $d$ affects the output as in Figure~\ref{feedblock}.
The system behavior is then described by \beqn y= Ae -d \qquad \mbox{where}
\qquad e = u - Fy. \eeqn Eliminating $e$, we find \beqn y = \frac{Au -
d}{1+AF}. \eeqn The sensitivity of the output to the disturbance is then
\beqn \frac{\partial y}{\partial d} = - \frac{1}{1+AF}. \eeqn Again, we see
that the sensitivity decreases as the loop gain $AF$ is increased.  Such a
disturbance could be caused by an increased load in a downstream part of an
electrical circuit or removal of end product from a biochemical pathway.

This property is of particular interest because the last stage of
MAPK, that is MAPK-PP (see Figure 2) has to migrate to the nucleus
in order to elicit a response. This diffusion is effectively a load
on the MAPK circuit. Without feedback such a load would have a
deleterious effect on the functioning of the MAPK pathway. As long
as there is a pool of unphosphorylated MAPK, the feedback is able to
compensate for this increased load. It is only when that pool dries
up (as the amplifier is saturated) that the feedback is unable to
provide any benefit. This property also has the natural benefit of
automatically modularizing the network into, effectively, a single
functional unit.

\paragraph{Improved linearity of response over extended operating range} Consider now the case
where the response of the amplifier is a nonlinear function of the input
signal, so the open loop response is $y=A(u)$, e.g.~$y=10 \frac{u}{1+u}$).
Taking the feedback again to act simply by multiplication, the behavior of
the system $G$ (now also a function of  $u$) is described by \beqn y = G(u)
=A(e) \qquad \mbox{where} \qquad e = u - Fy = u - FG(u). \eeqn
Differentiating we find \beqn G^\prime(u) = A^{\prime}(e)  \frac{de}{du} =
A^{\prime}(e) ( 1 - F G^\prime (u)). \eeqn Solving for $G^\prime(u)$ we
find \beqn G^\prime (u) = \frac{A^\prime(e)}{1 + A^\prime(e)F}. \eeqn We
find then, that if $A^\prime(e)F$ is large ($A^\prime(e)F \gg 1$), then
\beqn G^\prime (u) \approx \frac{1}{F}, \eeqn so, in particular, $G$ is
approximately linear, as its derivative is approximately constant. In this
case, the linear feedback compensates for the nonlinearities in the
amplifier $A(\cdot)$. Another feature of this analysis is that the slope of
$G(\cdot)$ is less than that of $A(\cdot)$, i.e.~the response is
``stretched out''.  For instance, if $A(\cdot)$ is saturated by inputs
above and below a certain ``active range'', then $G(\cdot)$ will exhibit
the same saturation, but with a broader active range.

A natural objection to the implementation of feedback as described above is
that the system sensitivity is not actually reduced, but rather is shifted
so that the response is more sensitive to the feedback $F$ and less
sensitive to the amplifier $A$.  However, in each of the cases described
above, we see that it is the nature of the loop gain $AF$ (and not just the
feedback $F$) which determines the extent to which the feedback affects the
nature of the system.  This suggests an obvious strategy.  By designing a
system which has a small ``clean'' feedback gain and a large ``sloppy''
amplifier, one ensures that the loop gain is large and the behavior of the
system is satisfactory.  Engineers employ precisely this strategy in the
design of electrical feedback amplifiers, regularly making use of
amplifiers with gains several orders of magnitude larger than the feedback
gain (and the gain of the resulting system).

In summary a feedback amplifier provides the following desirable
characteristics:

\begin{description}
\item 1. Increased robustness with respect to internal perturbations.
\item 2. Insulation from external perturbation, resulting in functional modularization.
\item 3. A linear graded response over an extended operating range.
\end{description}

It is hard to overstate the importance of the feedback amplifier
in circuit design. One of the most important components in analog
electronics is the operational amplifier, or op-amp. The original
concept of the op-amp came from the field of analog computers in
the 1940s where extremely high-gain DC amplifiers were employed
to carry out analog computations such as addition,
multiplication, differentiation and most important of all,
integration. The operating characteristics of a particular op-amp
were determined by the feedback elements. By changing the
arrangements and types of feedback elements different analog
operations could be implemented. Thus the same amplifier was able
to perform a variety of operations with only small changes to the
feedback elements.

One of the key characteristics of an op-amp is the extremely high
gain between the input and output signals. By analogy we can see
that the basic MAPK cascade pathway, without the feedback loop,
has the same high gain characteristic. In the next section we shall
illustrate how a high gain MAPK cascade with negative feedback will
act in an exactly analogous manner to the feedback amplifier.

\paragraph{Interconvertible Enzyme Cascades as Feedback Amplifiers}
Having described the utility of negative feedback in electrical
amplifiers, we now demonstrate that the same benefits can be reaped
when feedback is wrapped around a biochemical amplifier such as the
MAPK cascade.  Our analysis is based on a model from Ferrell and
Kholedenko~\citep{Ferrell:1996,BKoholodenko:2000}, with a negative
feedback of the form described in previous work incorporating
similar
mechanisms~\citep{Anad:2001,BKoholodenko:2000,Brightman:2000}. Note
that our conclusions do not depend explicitly on this particular
model but instead is a function of the high gain achieved by the
cascade of cycles and the presence of the negative feedback. The
model simply serves as a means to illustrate the concept. The
network is shown in Figure~\ref{network}. We begin by describing
some simulations which show that this nonlinear system exhibits
behavior similar to that described above for feedback amplifiers.

\begin{figure}
\begin{center}
  \includegraphics[scale=0.8]{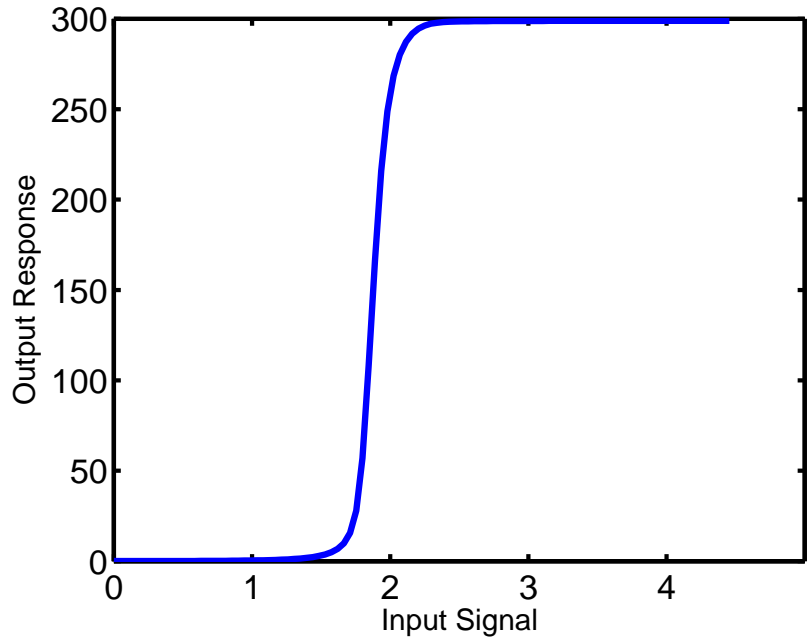}
\end{center}
\caption{Response of MAPK model in the case of no negative feedback}
\label{switch}
\end{figure}

At the basal level of no feedback, the steady state behavior of the
pathway is shown in Figure~\ref{switch}.  As discussed above, the
cascade acts as a switch -- the concentration of activated MAPK
climbs quickly from its basal level to saturation (i.e. complete
activation) at a particular ``threshold'' concentration of
signal~\citep{Goldbeter:1984,Ferrell:1996}. For inputs near the
threshold level, the pathway acts as an amplifier, with the output
level of activated MAPK far exceeding the signal concentration.
(This is a feature which is strongly dependent on the increasing
concentrations of kinases at successive levels of the
cascade~\citep{Goldbeter:1984,Anad:2001}.)

\begin{figure}
\begin{center}
  \includegraphics[width=6in,height=3.2in]{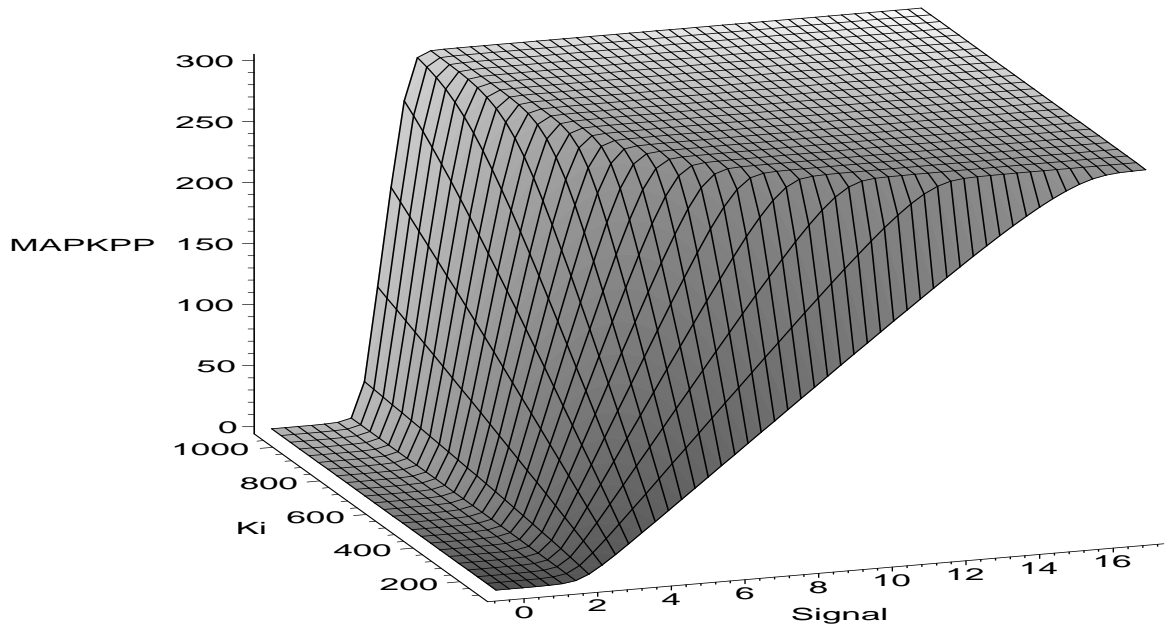}
\end{center}
\caption{Response for various levels of feedback gain}
\label{gradedfeedback}
\end{figure}

Figure~\ref{gradedfeedback} shows the input/output (Sig to MAPK-PP)
behavior of the system as the feedback gain (1/Ki) is increased.
Here we see, as noted in~\citep{BhallaRam} the `S'-shaped response
curve is stretched and straightened.  As the feedback is increased,
the operating range of the amplifier is increased and the response
is more linear over that range.  It is also clear that the overall
gain is decreasing, since the response for each particular input
level diminishes as the feedback gain increases.  It is argued
in~\citep{BhallaRam} that this behavior allows the cascades to be
bi-functional, acting as a switch when the feedback is low and as an
amplifier when the feedback is high.  From the point of view of the
amplifier characteristics, a switch can be thought of as an
amplifier with very short operating range.  An increase in the
feedback gain increases the range over which the cascade acts as an
amplifier.

\begin{figure}
\begin{center}
   \includegraphics[scale=0.51]{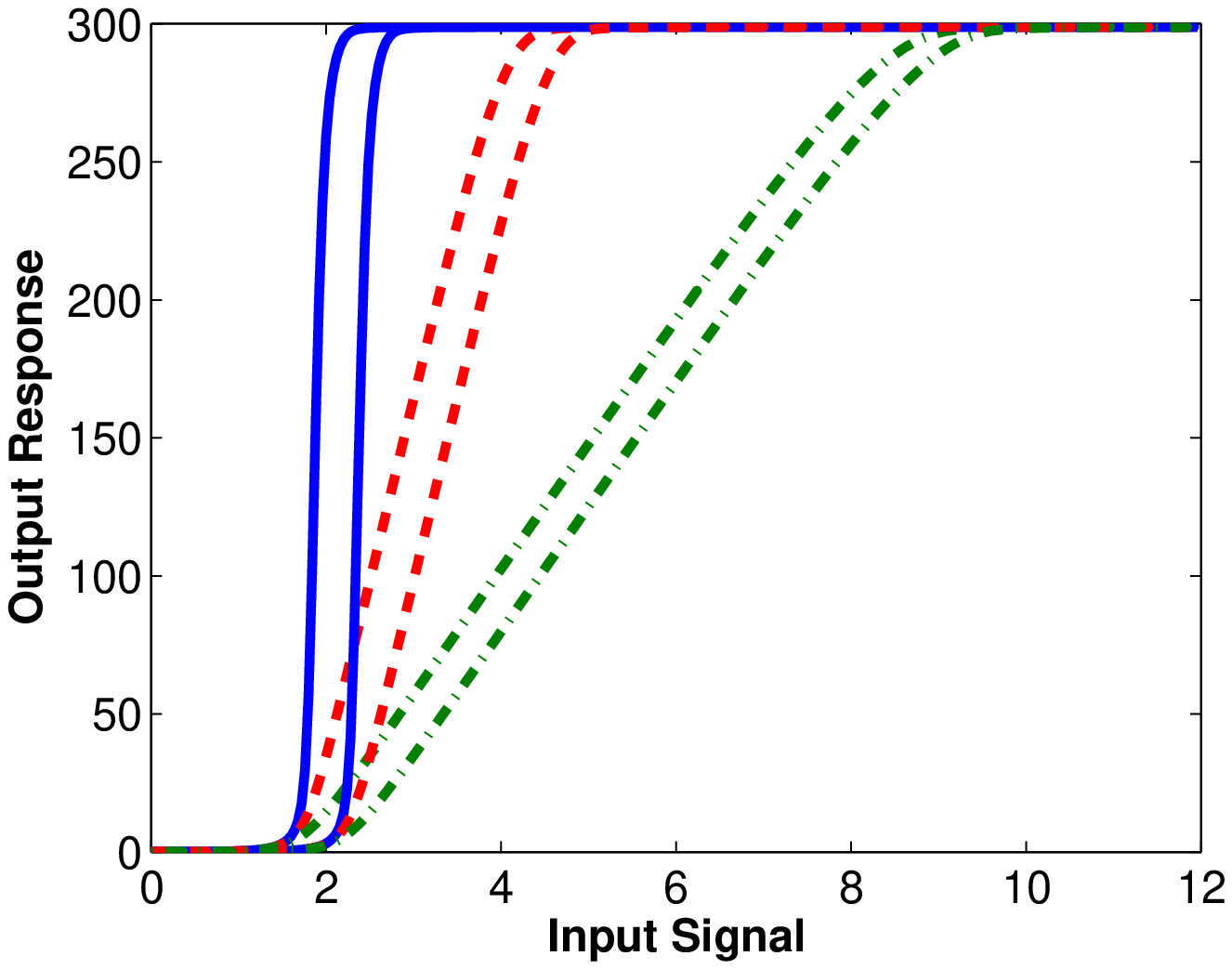}
   \includegraphics[scale=0.51]{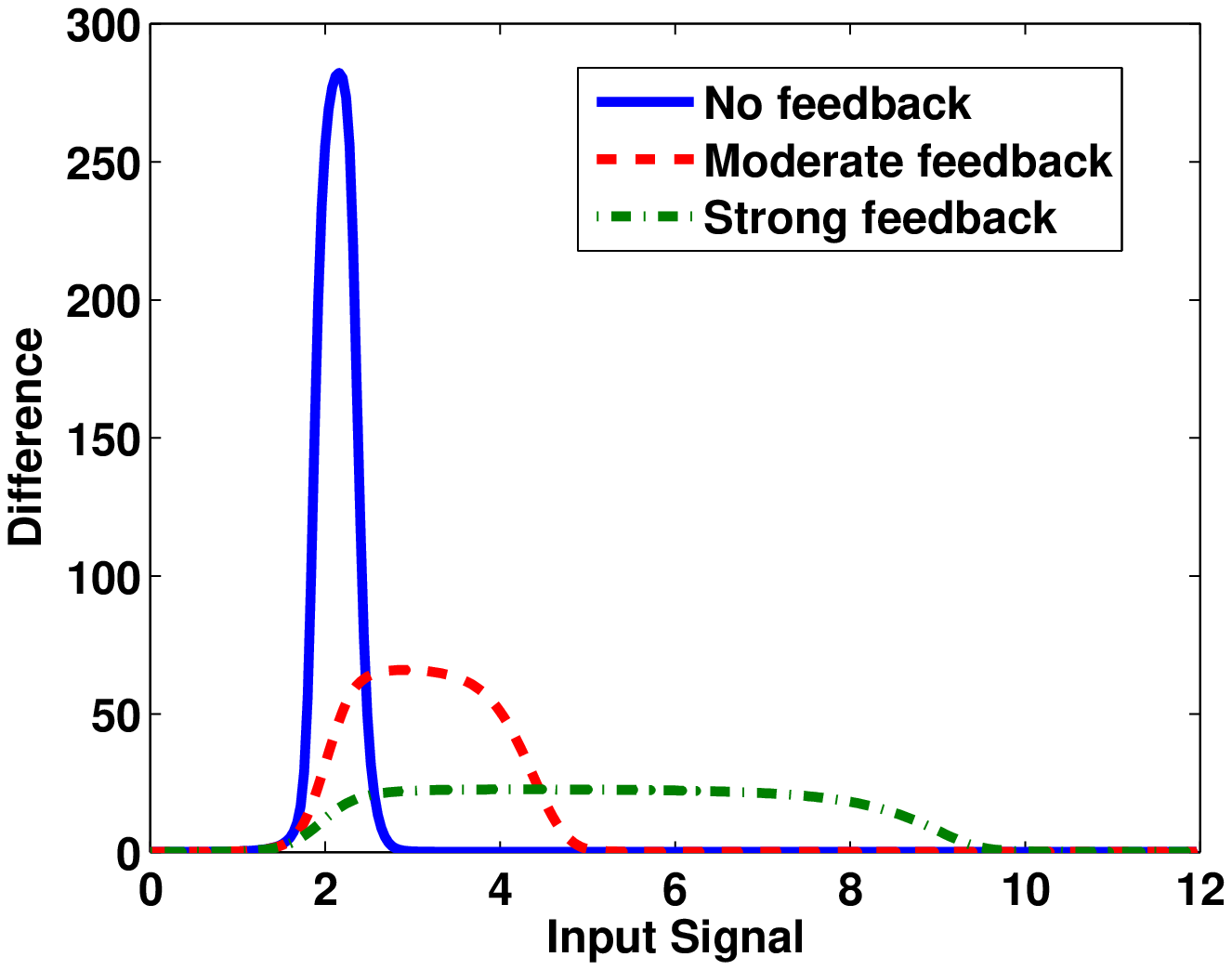}
\end{center}
\caption{Effect on response on a change in $V_1$ (See equation (4)). In the
first panel, pairs of graphs are shown, each pair represents a wild type
and a perturbation to $V_1$. Individual pairs are measured under differing
feedback strengths (Solid: no feedback (Ki = 10000), Dash: moderate level
of feedback (Ki = 600), Dash-Dot:strong feedback (Ki = 200)). The second
panel shows the difference in response between each pair. Note that as the
feedback increases, the difference becomes more linear. } \label{intpara}
\end{figure}

Figure~\ref{intpara} shows the effect of parameter variation on the
system at various levels of feedback intensity.  The response of the
system is plotted against signal level for three different levels of
feedback gain.  In each case, the response is also shown after a
perturbation in the kinetic parameter $V_1$ (a 20\% decrease).  Also
shown is a plot of the {\em difference} in response between the
nominal and perturbed systems.  The feedback is again having a
``stretching'' effect.  While the ``total'' error is not appreciably
diminished, the effect of the feedback is to spread that error out
across the increased operating range of the amplifier.  The result is
a significant decrease in the change in response for each particular
input level within the active range.  Similar result hold for
perturbations in the other model parameters.



\paragraph{Linearized model} In addition to simulations of the complete
model, analysis of a linearization is also useful in drawing analogy to the
linear feedback amplifier described above.  The model was linearized about
a nominal operating point (steady state) with moderate feedback gain
($\mKi=600$) and an input level in the operating range of the amplifier
($\mbox{Sig}=3$). The linearization takes the form

\beqn \dot{x} = A x + B u - F(\mKi) x \eeqn

where the state variables $x$ and the input $u$ describe the deviation of
the enzyme concentrations and signal level from their nominal values.
This is a slightly different form than the linear amplifier discussed above
since the feedback acts on the system rather than directly on the input
(i.e. the feedback influences the effect of \Sig~ on \KKK~ rather than
having a direct effect on \Sig~ itself).  In this case we expect the closed
loop gain to have the form \beqn y = \frac{A_1 u}{1 + A_2 F} \eeqn where
$A_1$ is the gain from input to \KKKP, and $A_2$ is the gain from \KKKP~ to
\KPP.  In this case the feedback is described as \beqn F(\mKi) =
\frac{4.87}{(1 + \frac{164}{\mKi})^2 \mKi} \eeqn so that \beq
\label{feednum} \frac{A_1 u}{1 + A_2 F(\mKi)} = A_1 u \frac{ \mKi^2 + 328
\mKi + 26900}{ \mKi^2 + (328+4.87A_2) \mKi + 26900} \eeq The linear (steady
state) relationship between input and output (i.e.~signal level and
MAPK-PP) is given by \beqn \mKPP - \mKPP_0 = (\mSig - \mSig_0) 2180 \frac{
\mKi^2 + 329 \mKi + 27000}{ \mKi^2 + 15600 \mKi + 27000}, \eeqn where
$\mKPP_0$ and $\mSig_0$ are the nominal levels. This matches~\rref{feednum}
with $A_1 = 2180$, $A_2 = 3140$.

\begin{figure}
\begin{center}
  \includegraphics[scale=0.8]{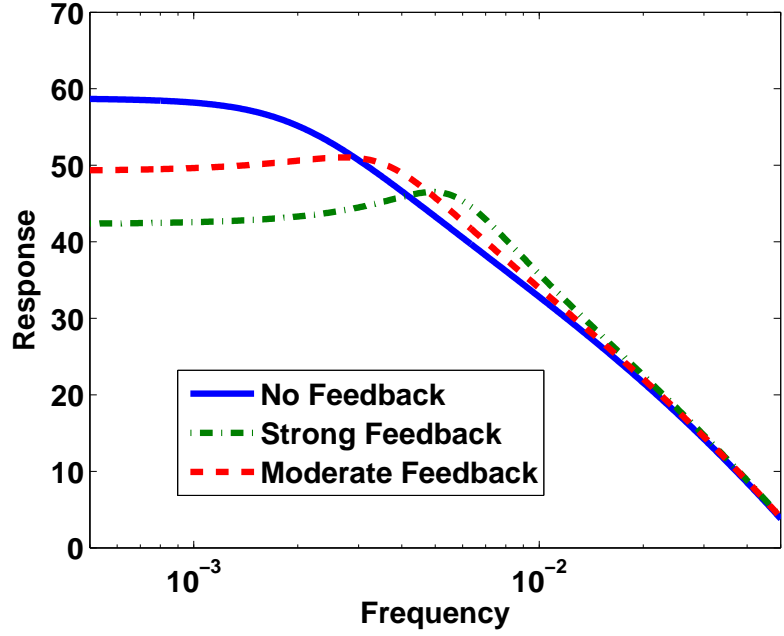}
\end{center}
\caption{Frequency response of linearized model for various feedback gains}
\label{bode}
\end{figure}

Another way of characterizing the linear system is through its
frequency response.  Graphs of the response versus frequency
(i.e.~magnitude Bode plots) are shown in Figure~\ref{bode} for various
levels of feedback.  This is the familiar response of a feedback
amplifier -- as the feedback increases, the gain of the system
decreases, but the range of frequencies over which that gain is
maintained is increased.  This highlights another feature of adding
negative feedback to an amplifier: an increase not just in the range of
constant signals that are amplified, but a corresponding increase in
the range of sinusoidal signals that are amplified.  Although this
amplification of oscillatory signals has not been seen as biologically
significant, one cannot rule out that some cells may be using the MAPK
cascade for this purpose, especially in light of the number
of oscillatory signals which have been identified.

\paragraph{Computational Model} The model network is shown in
Figure~\ref{network}. The eight species concentrations of interest are
\KKK, \KKKP, \KK, \KKP, \KKPP, \K, \KP, \KPP.  Moiety conservation allows
us to treat three of these as dependent variables through \beqn
\mKKK &=& \mbox{T}_1 - \mKKKP \\
\mKK &=& \mbox{T}_2 - \mKKP - \mKKPP \\
\mK & = & \mbox{T}_3 - \mKP - \mKPP. \eeqn The reactions are described by
Michaelis-Menten rate laws:

\begin{eqnarray}
      v_1 &=&  \frac{(\mV_1+ \mSig) \mKKK}{ (\mKm_1 + \mKKK) \left (1 +
\frac{\mKPP}{\mKi} \right )}\\ \nonumber
      v_2 &=&   \frac{\mV_2 \, \mKKKP}{\mKm_2 + \mKKKP}\\ \nonumber
      \\ \nonumber
      v_3 &=&    \frac{\mV_3 \, \mKK \, \mKKKP}{\mKm_3 + \mKK}\\  \nonumber
      v_4 &=&   \frac{  \mV_4 \, \mKKP}{\mKm_6 + \mKKP}\\  \nonumber
      v_5 &=&  \frac{\mV_5 \, \mKKP \, \mKKKP}{\mKm_5 + \mKKP}\\  \nonumber
      v_6 &=&  \frac{\mV_6 \, \mKKPP}{\mKm_4 + \mKKPP}\\ \nonumber
      \\ \nonumber
      v_7 &=&       \frac{\mV_7 \, \mK \, \mKKPP}{\mKm_7 + \mK}\\ \nonumber
      v_8 &=&      \frac{ \mV_8 \, \mKP}{\mKm_{10} + \mKP}\\ \nonumber
      v_9 &=&    \frac{\mV_9 \, \mKP \, \mKKPP}{\mKm_8 + \mKP}\\ \nonumber
      v_{10} &=&   \frac{\mV_{10} \, \mKPP}{\mKm_9 + \mKPP}\\ \nonumber
\end{eqnarray}

Nominal parameter values are chosen as \beqn &&\mKm_1 = 10 \ \  \mKm_2 = 8
\ \ \mKm_3 = 15 \ \  \mKm_4 = 15 \ \ \mKm_5 = 15 \\&& \mKm_6 = 15 \ \
\mKm_7 = 15 \ \ \mKm_8 = 15 \ \ \mKm_9 = 15 \ \  \mKm_{10} = 15 \\&& \mV_1
= 2.5 \ \   \mV_2 = 5.1 \ \ \mV_3 = 0.025 \ \ \mV_4 = 0.75 \ \ \mV_5 =
0.025 \\&& \mV_6 = 0.75  \ \ \mV_7 = 0.025 \ \ \mV_8 = 0.5 \ \ \mV_9 =
0.025 \ \ \mV_{10} = 0.5  \\&& \mbox{T}_1= 100 \ \ \mbox{T}_2=300 \ \
\mbox{T}_3=300 \eeqn

The units for concentration and Michaelis constants are given in nM,
maximum rates are expressed as nM $s^{-1}$ and rate constants in $s^{-1}$.
To investigate the behavior of the system under a load on the output,
sequestration of \KKPP in the nucleus is included in the model by the
inclusion of an additional state variable nucMAPK-PP, described by \beqn
&&\mK  =  \mbox{T}_3 - \mKP - \mKPP - \mbox{nucMAPK-PP} \\  \\
   &&   v_{11} = k_{11} \mKPP \ \
      v_{12} = k_{12} \mbox{[nucMAPK-PP]} \\
& & k_{11} = 0.003 \ \ k_{12} = 0.01 \eeqn

The model was linearized about a nominal steady state with moderate
feedback level ($\mKi=600$) and a signal level in the amplifier range
($\mbox{Sig} = 3$).  The nominal values of the state variables are \beqn
&&\mKKKP_0 = 23.8 \ \ \mKKP_0 = 43.9 \ \
\mKKPP_0 = 21.7 \\
&&\mKP_0 = 82.3 \ \ \mKPP_0 = 164 \ \ \eeqn The linearized model takes the
form \beqn \dot{x} = A x + B u - F(\mKi) x \eeqn where
\begin{eqnarray}
&&x= \left ( \begin{array}{c} \mKKKP - \mKKKP_0 \\ \mKKP - \mKKP_0 \\
\mKKPP - \mKKPP_0 \\ \mKP - \mKP_0 \\ \mKPP - \mKPP_0 \\ \end{array} \right
) \qquad u = \mSig -\mSig_0  \nonumber \\
A &=&  \left ( \begin{array}{ccccc} -0.0462 & 0 &  0 & 0 & 0 \\
0.00487 &  -0.00596 &  0.00822 &  0 & 0 \\
0.0186 &  0.00257 &  -0.00837 & 0 & 0 \\
0 & 0 & -0.00163 &  -0.00339 &  -0.00151 \\ \nonumber
0 & 0 & 0.0211 & 0.000859 &  -0.000233 \end{array} \right ) \\
B &=& \left ( \begin{array}{c} 0.694 \\ 0 \\ 0 \\ 0 \\ 0 \end{array}
\right) \nonumber
\\  F(\mKi) &=& \left ( \begin{array}{ccccc} 0 & 0
&  0 & 0 & \frac{4.87}{(1 + \frac{164}{\mKi})^2 \mKi} \\
0 &  0 &  0 &  0 & 0 \\
0 &  0 &  0 & 0 & 0 \\
0 & 0 & 0 &  0 &  0 \\
0 & 0 & 0 & 0 &  0 \end{array} \right ) \nonumber
\end{eqnarray}

\section*{Discussion}

In this paper we have discussed an alternative hypothesis on the
functional role of the MAPK pathway. In particular, we have
discussed the possibility that MAPK is acting as a classic
feedback amplifier so beloved of engineers.

\paragraph{Experimental Evidence} Although the previous theoretical and
computational arguments offer a novel view of how the MAPK pathway
may be operating, the popularly held view is that MAPK pathways are
in fact operating as switch devices rather than generating a graded
response as we have suggested here. Indeed in some cases a switch
response makes a very satisfactory explanation, for example, oocyte
maturation presumably requires an all-or-none response to
progesterone~\citep{Ferrell:1996}. However in other cases, the
situation is not so easily decided. Hazzaline and Mahadevan
\citep{HazzalinGraded} have discussed at length the case for a
graded response over a switch like response. They provide evidence
for the graded regulation of immediate-early genes in cell culture
which are controlled via MAPK pathways. Furthermore, there is now
clear evidence to support graded responses in the yeast mating
pheromone pathway \citep{YeastGraded,Ferrell:2002} which is again
based on a MAPK cascade.

There are also a growing number of papers which report a graded
response of ERK (Mitogen activated protein kinase) with respect to
EGF (Epidermal Growth Factor) stimulation. The first paper
\citep{BhallaRam} describes a cell population study where the
response was investigated using Platelet-Derived Growth Factor
(PDGF), they observed a proportional response over a 10-fold
concentration range of PDGF. This effect was explained in terms of
the action of the negative feedback modulator, MKP (MAPK
Phosphatase). Although this feedback only operates on one level of
the three cascade structure, it still induces the property of
linearization due to negative feedback as described in this paper.

Much more convincing however is recent work carried out on single
cell studies. It is becoming increasingly apparent that population
level studies give only a rough indication of cellular dynamics,
this is due to heterogeneity in the population. A stricking example
of this effect are the oscillations recently observed in p53 in
response to DNA damage \citep{Lahav2004}. In population studies the
oscillations appear damped (due to each oscillator being out of
phase), while in single cell studies the oscillations are of fixed
amplitude. It is therefore of interest to note two papers
\citep{Whitehurst,MacKeigan} which focus on single cell studies. In
both cases they observed a clear graded response of ERK stimulation
with respect to EGF. In the first paper \citep{Whitehurst} the
concentration of dually phosphorylated ERK in individual HeLa cells
was correlated with activation as a result of EGF and PMA
stimulation. In both cases a clear linear response was observed. The
second paper \citep{MacKeigan} investigated the response of Swiss
3T3 fibroblasts to EGF and PDGF. The work used a FACS (Fluorescence
Activated Cell Sorting) based approach to investigate the
distribution of activation in the cell population. If the cascade
operated in an ultrasensitivity mode then the authors expected to
see two cell populations corresponding to an off an on state.
Instead the experiments showed intermediate activation of ERK in
response to EGF.

Although these experiments indicate that ERK responds linearly to
stimulations they do not suggest how the linearity originates. We
postulate in this paper that such a response would originate from
ultrasensitivity coupled with negative feedback. To determine
whether this is the case further studies would need to be
undertaken, one possibility would be to remove the feedback loop and
to remeasure the response. Further evidence of the role of the
negative feedback could be obtained by investigating the robustness
of the response to changes in proteins levels inside the loop. If
the feedback loop is operating then such changes should have little
effect on the network response. While such experiments would not
prove conclusively that the cascade is acting as a feedback
amplifier, they would certainly lend weight to that alternative.

\paragraph{Network Versatility} The MAPK unit appears to be a very versatile
unit \citep{BhallaRam}. Without any feedback, MAPK can act as a very
high gain amplifier, so high that is behaves in a switch like manner
\citep{Ferrell:1996,BagowskiFerrell2003}. With the addition of
feedback a number of new behaviors can emerge. With positive
feedback, MAPK can act as a bistable switch, with negative feedback
MAPK can act as a classic feedback amplifier. Moreover, with
sufficient negative feedback the MAPK pathway can be made to
oscillate \citep{BKoholodenko:2000}. Many of these behaviors have
now been observed experimentally
\citep{Ferrell:1996,BhallaRam,Whitehurst,MacKeigan}.


One of the great challenges in modern molecular biology (or systems
biology) is to understand the functioning of large complex networks.
Such networks are composed of thousands of reactions and without
some way to simplify their description we will find them almost
impossible to understand. In electrical engineering, it is now
almost routine to design complex circuits composed of millions of
components, for example the latest Pentium 4 from Intel is
reportedly composed of forty three million transistors. How, one may
ask, do engineers cope with such enormous complexity? A large part
of the answer like in modularization, that is, the network is broken
down into smaller function units which have distinct and well
defined behaviors. These in turn are modularized further, as
appropriate, down to the level of the discrete transistor. Thus a
hierarchy of function is described. By modularizing, each module
becomes manageable in terms of design and understanding. In reaction
networks, we need to take the same approach. Although the concept of
modularization \citep{LauffenburgerModules} in cellular networks
has, in recent years, received much attention, most studies have
concentrated on topological modularization
\citep{Milo2002,Ravasz2002,Lee2002}.

However we have one major problem which the engineers do not have.
Whereas man-made devices are designed, natural systems are clearly
not. Thus we need to be able to reverse engineer the modules. This
is obviously no easy task, for example we do not even know whether
natural selection would automatically generate a module based
structure. Experiments on artificially evolved systems
\citep{KozaBookIII,Deckard2004} suggest that both modular and non-modular
`designs' can evolve, with modularity being the most common outcome.
If this is the case for natural systems then we will be able to make
progress.

In this paper we describe the properties of one possible module,
that is the MAPK module. Many other modules most probably exist. We
are currently investigating other potentially functional units from
both prokaryotic and eukaryotic systems. Others have already started
using this approach \citep{Smolen1998,SmolenReview,SmolenLongReview}
and two excellent reviews in this area by Wolf and Arkin
\citep{WolfArkin2003} and Tyson et.\ al.\ \citep{TysonCell2003} are
well worth reading. This area of research would probably gain much
from looking at the sorts of devices that electronic engineers
employ, including devices such as switches, amplifiers, oscillators,
frequency filters amplitude filters, noise filters and amplifiers,
combinatorial logic, homeostats, rheostats, logic gates and memory
elements (list taken from \citep{WolfArkin2003}). For Systems
Biology to succeed, knowledge and experience from different
disciplines clearly needs to come together and electrical
engineering has probably a significant contribution to make in
allowing us to understand complex biological networks.

\section*{Materials and Methods}

All simulations were carried out using a combination of Matlab
(\url{http://www.mathworks.com/}) and the Systems Biology
Workbench~\citep{Sauro:Omics}. Graphs were generated using Matlab.

\section*{Acknowledgements}

We are grateful to John Doyle, Hana El-Samad and Mustafa Khammash
for very useful discussions. HMS acknowledges the support of the
Japan Science and Technology Corporation's ERATO Kitano Systems
Biology Project; BI acknowledges Air Force Research Laboratory
Cooperative Agreement No. F30602-01-2-558. We would also like to
acknowledge Prahlad Ram for bringing to our attention their paper.

\section*{Author Contributions}

The work described in this paper was carried out equally by both
authors, HMS and BI.

\bibliography{mybib}
\bibliographystyle{jtb}

\end{document}